\documentclass[iop]{emulateapj}

\usepackage{amsmath,amssymb,bm}
\usepackage[colorlinks]{hyperref}
\usepackage[all]{hypcap}
\usepackage{graphicx,float,epsfig}
\usepackage{mathrsfs,wasysym}
\usepackage{epstopdf}
\usepackage{comment}
\usepackage{pbox}
\usepackage{array}
\usepackage{xspace}
\usepackage[usenames,dvipsnames]{color}

\DeclareMathAlphabet\mathbfcal{OMS}{cmsy}{b}{n}

\DeclareSymbolFontAlphabet{\mathrsfs}{rsfs}
\DeclareMathAlphabet{\mathcal}{OMS}{cmsy}{m}{n}

\DeclareSymbolFont{bbold}{U}{bbold}{m}{n}
\DeclareSymbolFontAlphabet{\mathbbold}{bbold}

\newcommand{\be}{\begin{equation}}
\newcommand{\ee}{\end{equation}}

\def\tomega{{\tilde{\omega}}}

\def\go{\mathrel{\raise.3ex\hbox{$>$}\mkern-14mu
             \lower0.6ex\hbox{$\sim$}}}
\def\lo{\mathrel{\raise.3ex\hbox{$<$}\mkern-14mu
             \lower0.6ex\hbox{$\sim$}}}

\graphicspath{{figs/}}

\begin{document}

\title{Self-Trapping of Diskoseismic Corrugation Modes in Neutron Star Spacetimes}

\author{David Tsang\altaffilmark{1}, George Pappas\altaffilmark{2}}
\altaffiltext{1}{Center for Theory and Computation, Department of Astronomy, University of Maryland, College Park, MD 20742, USA }
\altaffiltext{2}{Department of Physics and Astronomy, The University of Mississippi, University, MS 38677, USA}

\begin{abstract}
We examine the effects of higher-order multipole contributions of rotating neutron star (NS) spacetimes on the propagation of corrugation (c-)modes within a thin accretion disk. We find that the Lense-Thirring precession frequency, which determines the propagation region of the low-frequency fundamental corrugation modes, can experience a turnover allowing for c-modes to become self-trapped for sufficiently high dimensionless spin $j$ and quadrupole rotational deformability $\alpha$. If such self-trapping c-modes can be detected, e.g. through phase-resolved spectroscopy of the iron line for a high-spin low-mass accreting neutron star, this could potentially constrain the spin-induced NS quadrupole and the NS equation of state.   
\end{abstract}

\section{Introduction}

The stationary axisymmetric Kerr spacetime can be characterized by only central object mass, $M$, and dimensionless spin, $j \equiv J/M^2$, i.e., the angular momentum of the central object over the square of its mass. The Kerr orbital and epicyclic frequencies for free particles near the equatorial plane are given by \citep{Okazaki1987, Kato1990},
\begin{align}
\Omega &= \frac{1}{M(r_{\rm BL}^{3/2} + j)},\\
\Omega_\perp &= \Omega\left( 1 - \frac{4j}{r_{\rm BL}^{3/2}} + \frac{3j^2}{r_{\rm BL}^2}\right)^{1/2},\\
\kappa &= \Omega\left( 1 - \frac{6}{r_{\rm BL}} + \frac{8j}{r_{\rm BL}^{3/2}} - \frac{3j^2}{r_{\rm BL}^2}\right)^{1/2},
\end{align}
where $\Omega$ is the orbital frequency, $\Omega_\perp$ and $\kappa$ are the vertical and radial epicyclic frequencies and $r_{\rm BL}$ is the Boyer-Lindquist radial coordinate divided by the mass.

The inequality between the orbital and vertical-epicyclic frequencies results in Lense-Thirring nodal-precession \citep{Bardeen1975}, and allows low-frequency corrugation waves to propagate within thin accretion disks, in the region interior to the inner-vertical resonance, where the wave frequency equals the Lense-Thirring precession frequency, $\Omega_{\rm LT}\equiv \Omega - \Omega_\perp$ \citep[see e.g.][]{Kato1990, SWO, Tsang2009a}. 

Trapped corrugation waves (c-modes) in thin relativistic disks have frequencies lower than associated inertial-acoustic (p-modes) and inertial modes (g-modes). They have been proposed as an explanation for low-frequency variability for accreting black hole (BH) systems. However, in order for c-modes to exist, a reflecting inner boundary must be present due to, e.g., a strong pressure maximum \citep{Kato1990}, or a truncation of the disk before the innermost stable circular orbit (ISCO), which will quickly damp modes unless the density transition towards the disk sonic point is extremely sharp \citep{Lai2009, Miranda2015}. 

Due to the theoretical uncertainty associated with such reflecting inner disk boundaries, much of the work in relativistic diskoseismology has focussed on g-modes, which for \citep[non-magnetized, see][]{Fu2009} thin BH disks can become self-trapping due to the general-relativistic turnover of the radial epicyclic frequency, independent of the inner disk boundary condition \citep[see][for reviews]{Wagoner1999, Kato2001, Lai2013}. 

In this letter we examine the trapping of disk c-modes in rotating neutron star (NS) spacetimes, that can significantly differ from Kerr in their higher-order multipole moments. We will show that for a certain range of NS parameters, the modification of the orbital and epicyclic frequencies due to the effect of a large spacetime quadrupole \citep{Pappas2012a,Gondeketal,Wisniewiczetal,Pappas2015} can result in the self-trapping of c-modes due to general-relativistic effects in the disk exterior to the ISCO and the stellar surface. The effects of the quadrupole on geodesics have been of interest to several authors \citep[see e.g.][]{Morsink1999ApJ,Pachon2012ApJ,Baubock2012ApJ,Baubock2014,Quevedo2015}. The observational identification of such a self-trapping c-mode in a spun-up accreting neutron star system through (e.g.) phase-resolved iron-line spectroscopy \citep{Tsang2013} could allow the quadrupole moment of the NS to be constrained.  

\section{Characteristic Frequencies of Neutron Star Spacetimes}
Near NSs additional parameters beyond the mass and spin are needed to characterize spacetime metric \citep[see e.g.][]{Pappas2009,Pappas2013}. Here we expand the appropriate free-particle orbital and epicyclic frequencies to include up to the mass-quadrupole, $M_2$, and spin-octupole, $S_3$. We restrict our interest to the characteristic frequencies for particles in the equatorial plane, as the high-spin NSs almost certainly have angular momentum dominated by accretion-induced spin-up. 

In a semi-Keplerian approach, these frequencies can be approximated as, 

\begin{align}
\Omega &= \frac{1}{M(r_{\rm circ}^{3/2} + j)}\left(1-\frac{3 \alpha  j^2}{4 r_{\rm circ}^2}+\frac{(1 -4 \alpha)  j^2}{2 r_{\rm circ}^3}\right.\nonumber\\
&\qquad\qquad\qquad\qquad~ \left.-\frac{3 \left(\alpha -2 \beta \right) j^3}{4 r_{\rm circ}^{7/2}}\right)^{-1}, \label{eq:NSOmega}\\
\Omega_\perp &= \Omega\left(1-\frac{4 j}{r_{\rm circ}^{3/2}} +\frac{3 \alpha  j^2}{r_{\rm circ}^2} -\frac{6 \left(1
   -\alpha \right) j^2}{ r_{\rm circ}^3}\right. \nonumber \\
   &\qquad ~~~~\left.-\frac{3 \left(4 \beta -3 \alpha \right) j^3}{ r_{\rm circ}^{7/2}}\right)^{1/2}, \label{eq:NSOmegaperp}\\
\kappa &=\Omega\left(1-\frac{6}{r_{\rm circ}}+\frac{8 j}{r_{\rm circ}^{3/2}}-\frac{3 \alpha  j^2}{r_{\rm circ}^2}\right.\nonumber\\
&\qquad~~~~\left.+\frac{3 \left(4 -5 \alpha  \right) j^2}{ r_{\rm circ}^3}+\frac{6 \left(3 \beta -2 \alpha \right) j^3}{r_{\rm circ}^{7/2}}\right)^{1/2},\label{eq:NSkappa}
\end{align}
where $r_{\rm circ} \equiv \sqrt{g_{\phi\phi}}/M$ is the circumferential radius divided by the mass, $\alpha\equiv-M_2/(j^2M^3)$ is the 
quadrupole rotational-deformability, and $\beta\equiv -S_3/(j^3 M^4)$ is the 
spin-octupole rotational-deformability. $M_2$ and $S_3$ are the relativistic multipole-moments (RMMs)
\citep[see][for review]{Quevedo1990} and are calculated for NSs as described by \citet{Pappas2012b}. 
The frequencies are constructed for a general stationary axisymmetric spacetime produced by a series expansion of an Ernst potential in terms of RMMs \citep[similar to ][but resummed and expressed in terms of $r_{\rm circ}$]{Ryan95}.
These expressions are exact including up to the leading $S_3$ contribution, which corresponds to ${\cal O}(r_{\rm circ}^{-5})$ for $\Omega$, and ${\cal O}(r_{\rm circ}^{-13/2})$ for $\Omega_\perp$ and $\kappa$, and give an accurate description of the frequencies for values of $j\lo 0.8$.  

For $\alpha=\beta=1$, the expressions \eqref{eq:NSOmega}-\eqref{eq:NSkappa} yield the Kerr frequencies\footnote{In the equatorial plane of the Kerr metric  $r_{\rm circ}^2 = r_{\rm BL}^2 + j^2 + 2j^2/r_{\rm BL}$. By construction, the Kerr limit of Equations \eqref{eq:NSOmega}-\eqref{eq:NSkappa}, where $\alpha = \beta = 1$, are also accurate up to the order of the leading $S_3$ contribution.} for a black hole with spin parameter $j$. Alternatively, we can describe NSs by assigning ``neutron star values" to the parameters $\alpha$ and $\beta$. 

It was recently shown by \citet{Pappas:2013naa,Stein2014ApJ,YagietalM4,yiloveq} that the RMM of NSs that obey realistic equations of state (EOSs) satisfy universal 3-hair relations between higher-order moments and the mass, the spin, and the quadrupole (the three hairs). Thus, the higher-order RMMs of NSs can be expressed in terms of the first three using those relations. For this study, the only higher-order moment of interest is $S_3$, which is parameterized by $\beta$. Using the relation between $S_3$ and $M_2$, $\beta$ can be expressed in terms of $\alpha$ as \citep{Pappas:2013naa},

\begin{align}
\beta = \left(-0.36 + 1.48 \left( \sqrt{\alpha}\right)^{0.65}\right)^3 ,\label{eq:NS3hair}
\end{align}
thus resulting in expressions for the characteristic NS frequencies that depend only on the parameters $j$, $\alpha$, and NS mass $M$. For NSs, the spin parameter is constrained to be $j \lo 0.7$ (the Keplerian breakup limit, which is well within the range of validity of Equations \eqref{eq:NSOmega}-\eqref{eq:NSkappa}), while the parameter $\alpha$ ranges from $\sim1.2$ up to $\sim10$. In Section \ref{sec:EOS} we will discuss possible ranges for these parameters along with constraints based on NS spin and mass for realistic EOSs.  

\section{Diskoseismic Oscillations}

Following \citet{Okazaki1987, Tsang2009a} we consider a thin isothermal accretion disk around a NS, utilizing isothermal Newtonian (flat space) expressions for the disk perturbations, while adopting the relativistic free-particle characteristic frequencies for the disk elements. While this mixed Newtonian/relativistic approach is formally inconsistent, it suffices to capture the qualitative effects of the spacetime on diskoseismic modes.  A  fully relativistic approach \citep[see e.g.][]{IL1992, SWO, Perez1997} can be used in future work for more detailed quantitive study of such modes.

For a thin isothermal disk with unperturbed velocity ${\bf u}_o = (0, r\Omega, 0)$ in cylindrical coordinates $(r, \phi, z)$, the unperturbed vertical density profile can be given as
(for small $z \ll r$) 
\begin{align}
\rho_o(r,z) = \frac{\Sigma(r)}{\sqrt{2\pi} H} \exp\left(-z^2/2H^2\right).
\end{align}
Here $\Sigma(r)$ is the (vertically integrated) surface density and
$H=c_s/\Omega_\perp$ is the vertical scale height, where $c_s \ll r\Omega$ is the
isothermal sound speed.

Perturbing the mass and momentum conservation equations gives
\begin{align}
\frac{\partial}{\partial t}\delta \rho + \nabla\cdot(\rho_o \delta
{\bf u} + {\bf u}_o \delta \rho) &= 0~,\\
\frac{\partial}{\partial t}\delta {\bf u} + ({\bf u}_o \cdot \nabla)
\delta {\bf u} + (\delta {\bf u} \cdot \nabla) {\bf u}_o &= -\nabla
\delta h~,
\end{align}
with the enthalpy perturbation $\delta h \equiv \delta P/\rho = c_s^2
\delta \rho/\rho$, where we assume that the perturbations are also
isothermal.
Following \citet{Tsang2009a} \citep[see also][]{Zhang2006} we assume perturbations of the form $\delta h, \delta P, \delta {\bf u}, \delta
\rho \propto \exp(im\phi - i\omega t) {\rm H}_n(z/H)$, where ${\rm H}_n(z/H)$ is the Hermite polynomial of order $n$, $m$ is the azimuthal wavenumber, and $\omega$ is the angular frequency of the perturbation. 
After some manipulation while neglecting terms proportional to $dH/dr \sim O(1/r)$, the perturbation equations can be combined to give the wave equation for disk enthalpy perturbations,
\begin{align}
&\frac{d^2}{dr^2}\delta h - \left(\frac{d}{dr}\ln \frac{D}{r\Sigma}
\right)\frac{d}{dr}\delta h \nonumber \\
&\qquad +
\left[\frac{2m\Omega}{r\tomega}\frac{d}{dr}\ln \frac{D}{\Omega\Sigma}
- \frac{m^2}{r^2} - \frac{D(\tomega^2 - n\Omega_\perp^2)}{c_s^2
\tomega^2}\right] \delta h = 0~, \label{mastereq} 
\end{align}
where $D \equiv \kappa^2 - \tomega^2$, and $\tomega \equiv \omega - m \Omega$.
Here, to capture the general-relativistic effects, we allow the characteristic frequencies to be given by Equations \eqref{eq:NSOmega}-\eqref{eq:NSkappa}, with coordinate mapping $r_{\rm circ} = r/M$. This is more self-consistent but quantitatively different from the mapping adopted by e.g. \citet{Kato1990} or \citet{Tsang2009a}, which take $r_{\rm BL} = r/M$, though the results are qualitatively similar for the Kerr limit.

\section{Self-Trapping of Corrugation Modes}

\begin{figure}
\centering
  \includegraphics[width=0.9\columnwidth]{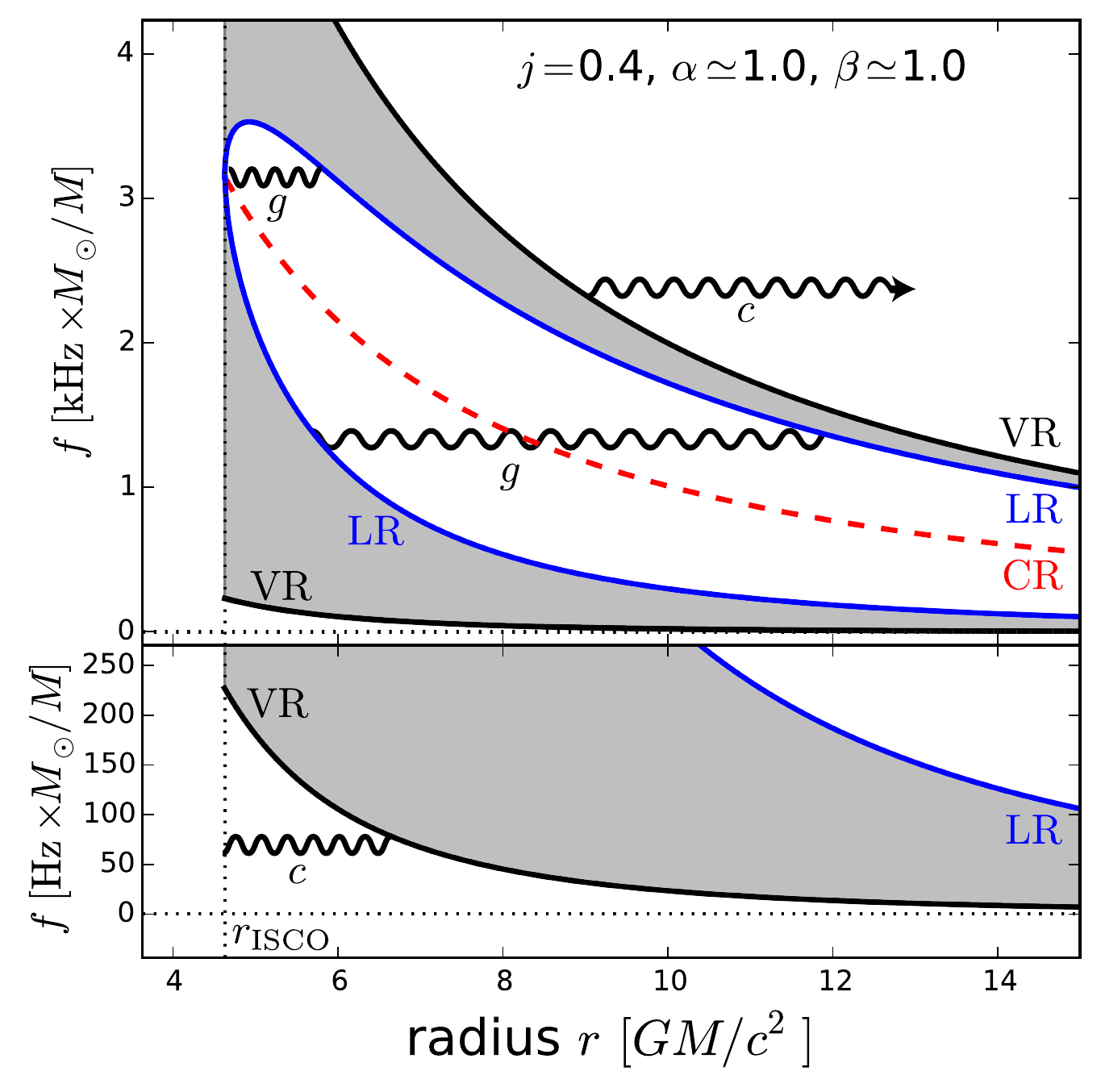}
  \caption{{\bf Upper}: Diskoseismic propagation diagram for a Kerr black hole ($\alpha=1$, $\beta = 1$) with spin parameter $j = 0.4$ for one-armed waves with $m=1, n=1$. Waves with frequency $f = \omega/2\pi$ can propagate in the white regions exterior to $r_{\rm ISCO}$, and are evanescent in the shaded regions between the vertical resonances (VR) and Lindblad resonances (LR). Inertial modes (g-modes), with $m=1, n=1$, can become self-trapping due to the turnover of the outer Lindblad resonance, while lower frequency g-modes are quickly damped by corotation (CR). Corrugation waves can propagate at high frequencies exterior to the outer VR, and at low-frequencies interior to the inner VR. {\bf Lower}: Enlargement of the propagation diagram at low frequencies. Corrugation modes (c-modes) can become trapped here if there exists a reflecting inner boundary at $r \go r_{\rm ISCO}$.
    }
  \label{fig:KerrProp}
\end{figure}

\begin{figure}
\centering
  \includegraphics[width=0.9\columnwidth]{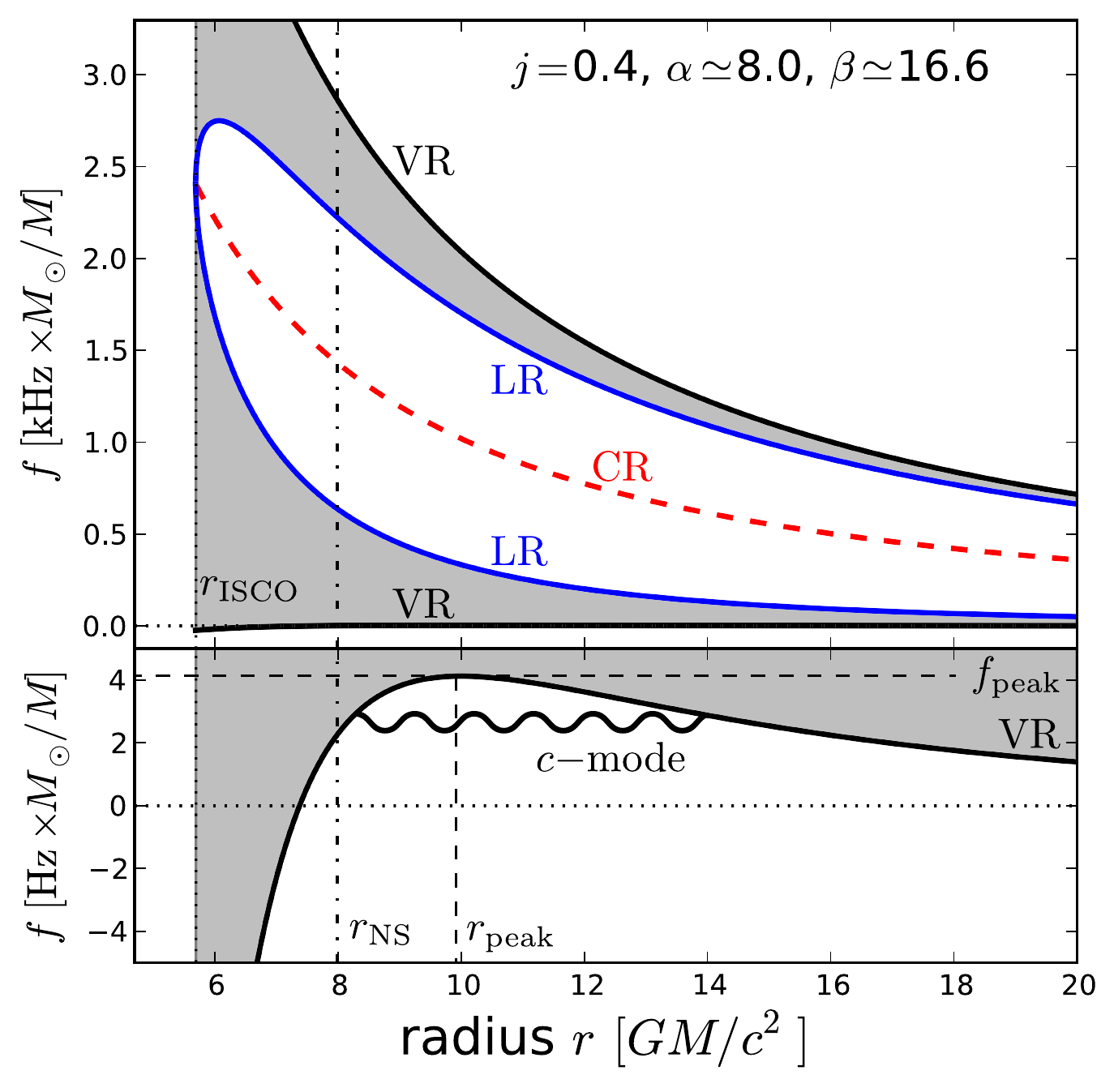}
  \caption{{\bf Upper}: Same as Figure \ref{fig:KerrProp} but for a neutron star spacetime with spin parameter $j = 0.4$, quadrupole rotational deformability $\alpha = 8$, and spin-octupole deformability $\beta \simeq 16.6$. Waves with frequency $f = \omega/2\pi$ can propagate in the white regions exterior to the NS radius $r_{\rm NS}$ (or wherever the disk is truncated).  Wave regions are qualitatively similar to the Kerr black hole (Figure \ref{fig:KerrProp}), except for the low-frequency c-mode region, where $\omega < \Omega - \Omega_\perp$. {\bf Lower}: At low frequencies c-modes can be self-trapped due to the turnover of the Lense-Thirring frequency, $\Omega - \Omega_\perp$, at radius $r_{\rm peak}$, and frequency $f_{\rm peak}$, as a result of the spacetime quadrupole contribution. 
    }
  \label{fig:NSprop}
\end{figure}

Far from the Lindblad resonances (where $D = 0$) and the corotation resonance (where $\tomega = 0$), we can adopt a WKB approach, assuming radial dependence $\delta h \propto \exp(i k r)$. This gives the WKB dispersion relation \citep{Okazaki1987}
\begin{align}
c_s^2 k^2= \frac{(\kappa^2 - \tomega^2)(n\Omega_\perp^2 - \tomega^2)}{\tomega^2},
\end{align}
where waves can propagate in regions where $k^2 > 0$. Modes with no vertical structure, such that $n=0$, are known as inertial-acoustic modes (p-modes) and propagate in the region where $\tomega > \kappa$ \citep[see e.g.][]{Lai2009, Tsang2009c}. Waves with $n > 0$ can propagate where $\tomega^2 < \kappa^2 < n\Omega_\perp^2$, the inertial-mode (g-mode) propagation region \citep[see e.g.][]{Wagoner2012}, or where $\tomega^2 > n\Omega_\perp^2 > \kappa^2$, the corrugation-mode (c-mode) propagation region. In this work we primarily consider the fundamental c-mode \citep{Kato1990}, with $m=1$, $n=1$, where the boundary of the propagation region is determined by the Lense-Thirring precession frequency $\omega = \Omega - \Omega_\perp$, though similar results will hold for any $m = \sqrt{n}$. 

%
\begin{figure}
\centering
  \includegraphics[width=0.9\columnwidth]{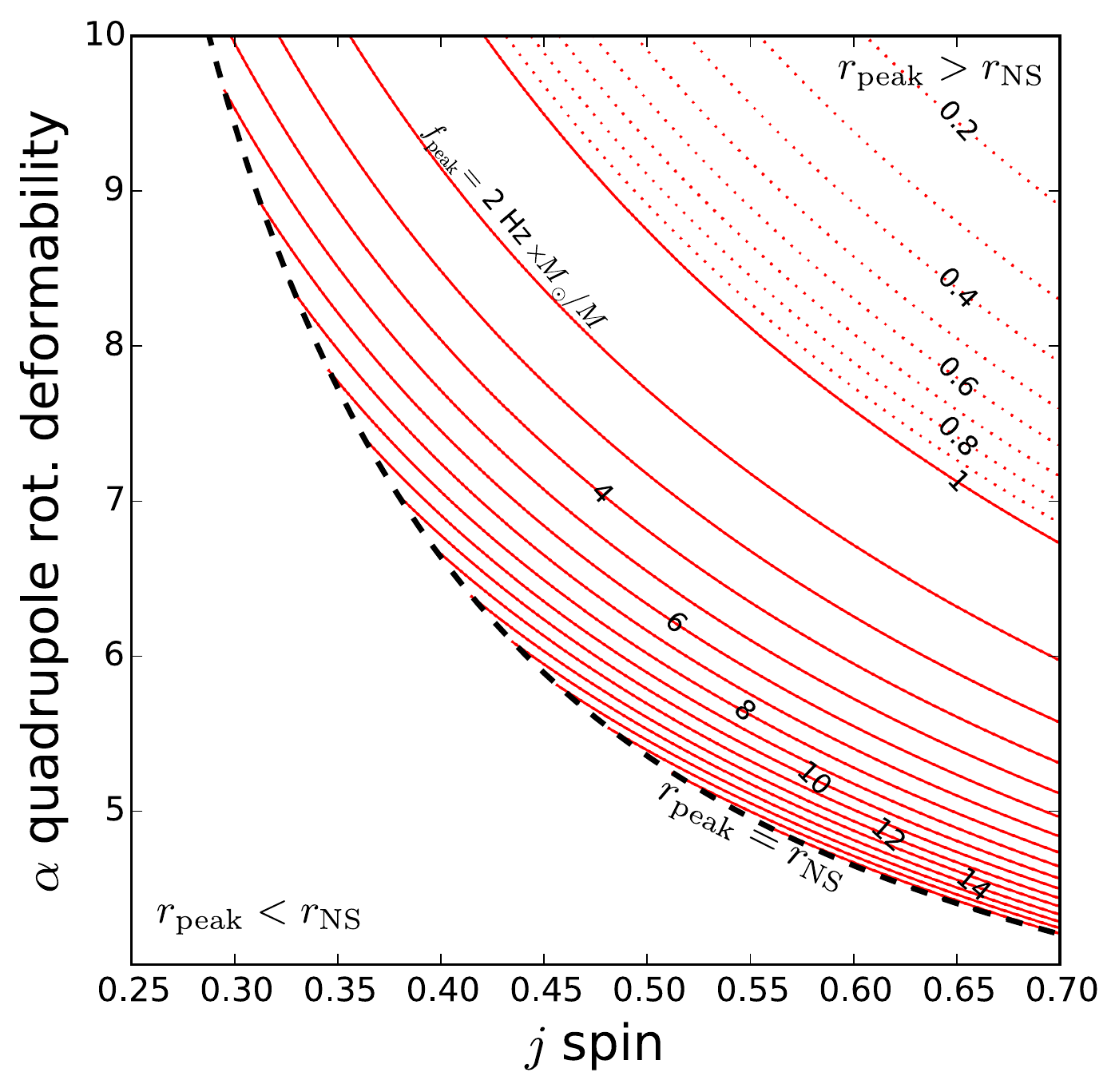}
  \caption{The contours of peak frequency, $f_{\rm peak}$ (in units of Hz$\times M_\odot/M$), at which the Lense-Thirring precession frequency $\Omega - \Omega_\perp$ turns over, as a function of spin $j$ and quadrupole rotational deformability $\alpha$. The dashed line indicates the parameter values for which the peak Lense-Thirring frequency occurs at a radius, $r_{\rm peak}$, coincident with the neutron star surface, $r_{\rm NS}$. For $j$ and $\alpha$ parameters above this line, the Lense-Thirring frequency peak occurs outside the star, and c-modes can become self-trapping.
    }
  \label{fig:LT_peakparam}
\end{figure}

In order for c-modes in the Kerr metric to be trapped a reflecting inner boundary is required such that a mode cavity can form (Figure \ref{fig:KerrProp}). 

For spacetimes with different RMM, the Lense-Thirring frequency can be significantly modified close to the central object. For some NS parameters this can result in a turnover of $\Omega_{\rm LT}$, allowing corrugation modes to become self-trapping, reflected on both sides by the inner vertical resonance (VR), where $\omega = \Omega_{\rm LT}$, as shown in the Figure \ref{fig:NSprop}. 

In Figure \ref{fig:LT_peakparam}, we show the peak of the Lense-Thirring frequency, $f_{\rm peak} = \Omega_{\rm LT}(r_{\rm peak})/2\pi$ (in units of ${\rm Hz} \times M_\odot/M$),  for the spacetime parameters $j$ and $\alpha$. We also indicate the $j$ and $\alpha$ where the radius at which $\Omega_{\rm LT}$ peaks, $r_{\rm peak}$, is equal to the expected equatorial radius of the neutron star, $r_{\rm NS}$. For NSs with realistic EOSs, 
the equatorial radius can be approximated by a single function of $j$ and $\alpha$ using the fitting formula\footnote{This fitting formula is accurate to within $\sim 6\%$ for $j \lo 0.65$.} from Appendix A of \citet{Pappas2015}.  

NSs with $j$ and $\alpha$ values above the line where $r_{\rm peak} = r_{\rm NS}$ have $r_{\rm peak}$ located outside the star, and thus can induce self-trapping of diskoseismic corrugation modes. The discrete c-mode spectra and detailed radial structure depend on the disk sound speed, such that a half-integer number of wavelengths fit in the cavity at the mode frequency. However, all self-trapped c-modes should have a radial trapping region with inner edge distinct from the disk edge, and frequency below $m\times f_{\rm peak}$ (see Figure \ref{fig:NSprop}), set by $m \times \Omega_{\rm LT}$ at both the inner and outer edges of the self-trapping region.

NSs with $j$ and $\alpha$ values at or below this line (or NSs that have disks magnetically truncated outside $r_{\rm peak}$), have maximum value of $\Omega_{\rm LT}$ occuring at $r_{\rm NS}$ (or the disk inner edge). In these systems c-modes must be trapped by reflection at the disk edge, with frequency set by $m\times \Omega_{\rm LT}(r_{\rm IVR})$ at the \emph{outer} edge of the trapping region, where the spacetime quadrupole effects are more subtle.

\begin{figure}
\begin{center}$
\begin{array}{ll}
\includegraphics[width=0.49\columnwidth]{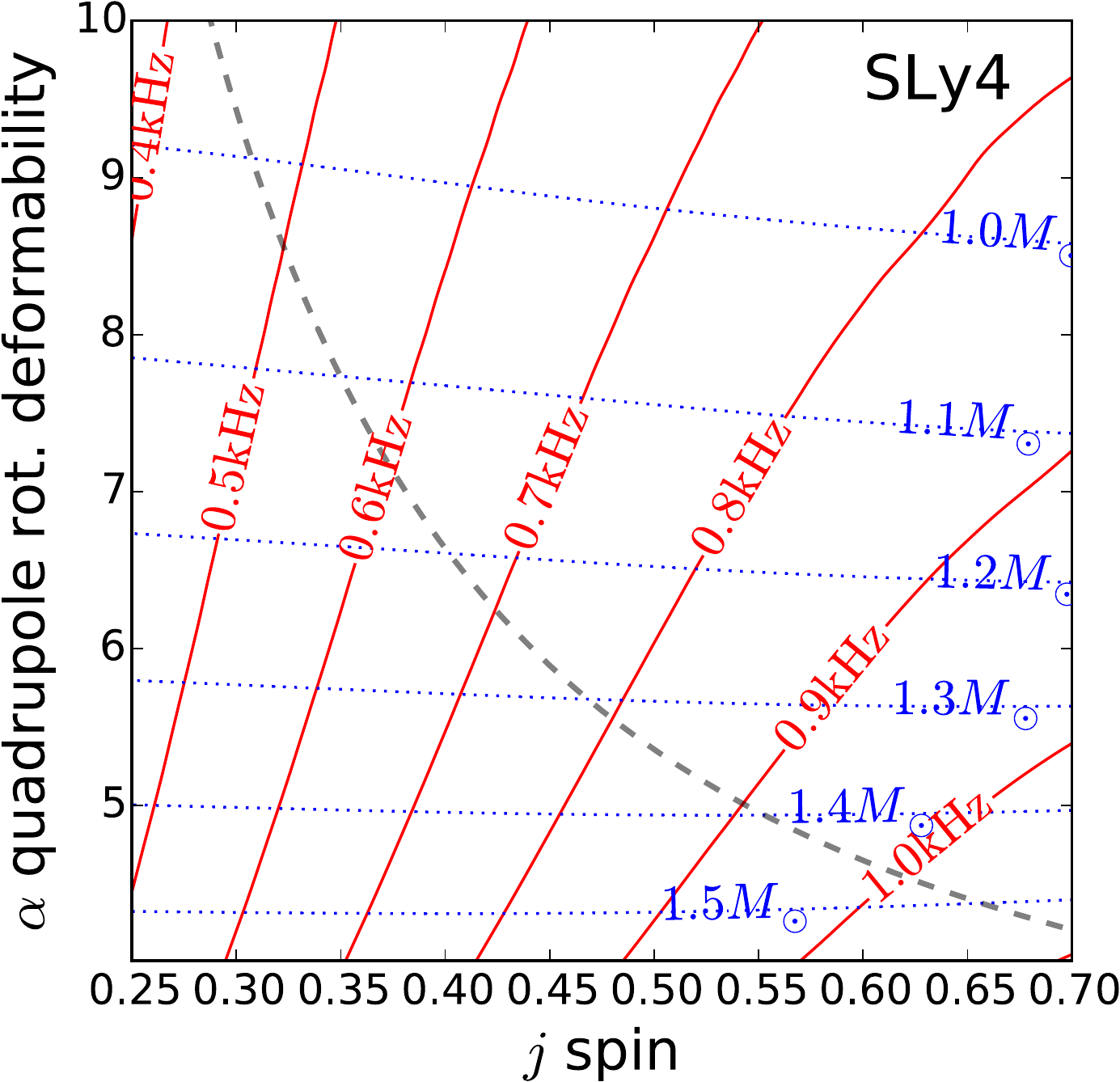} &
\includegraphics[width=0.49\columnwidth]{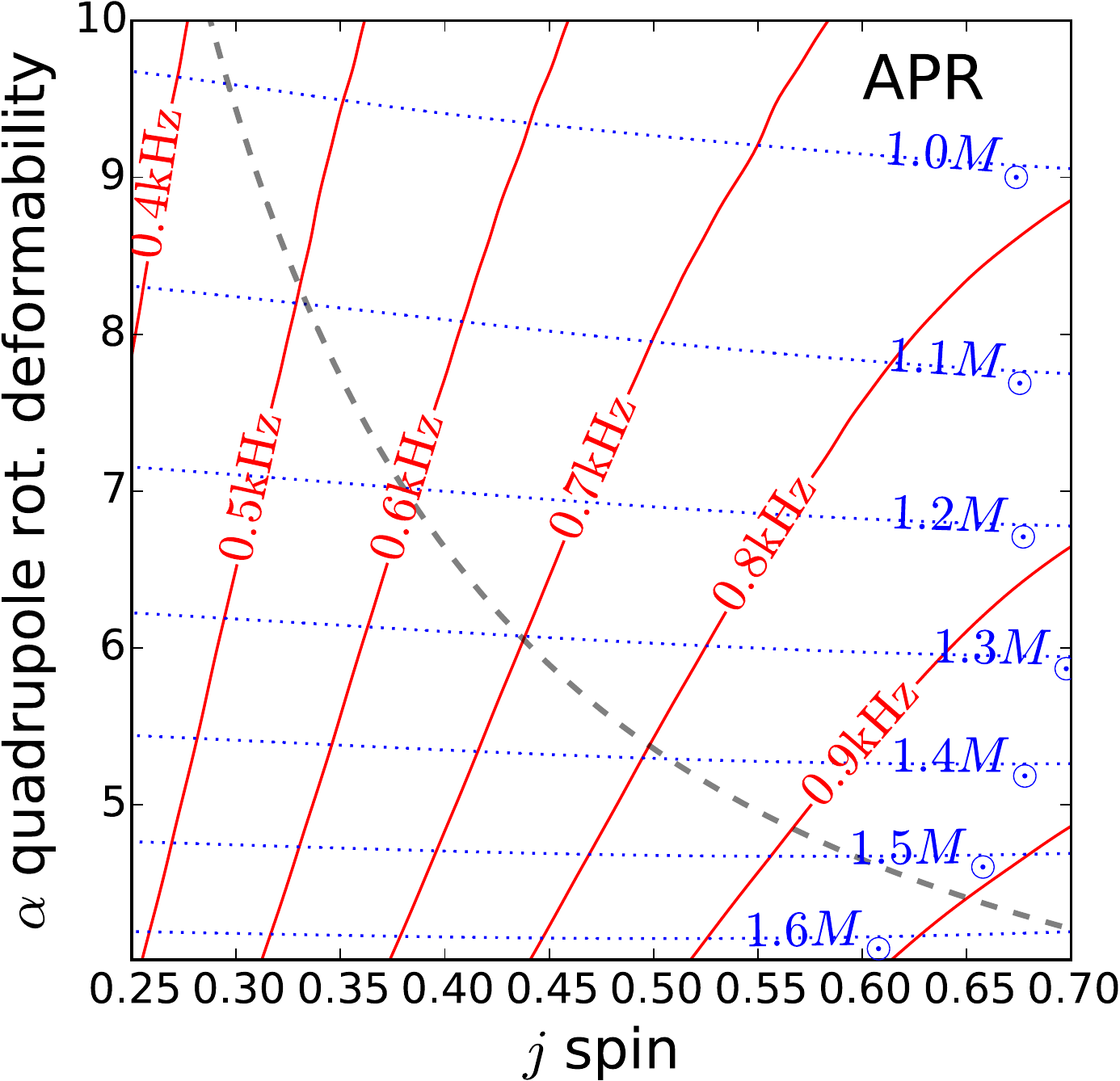}\\
\includegraphics[width=0.49\columnwidth]{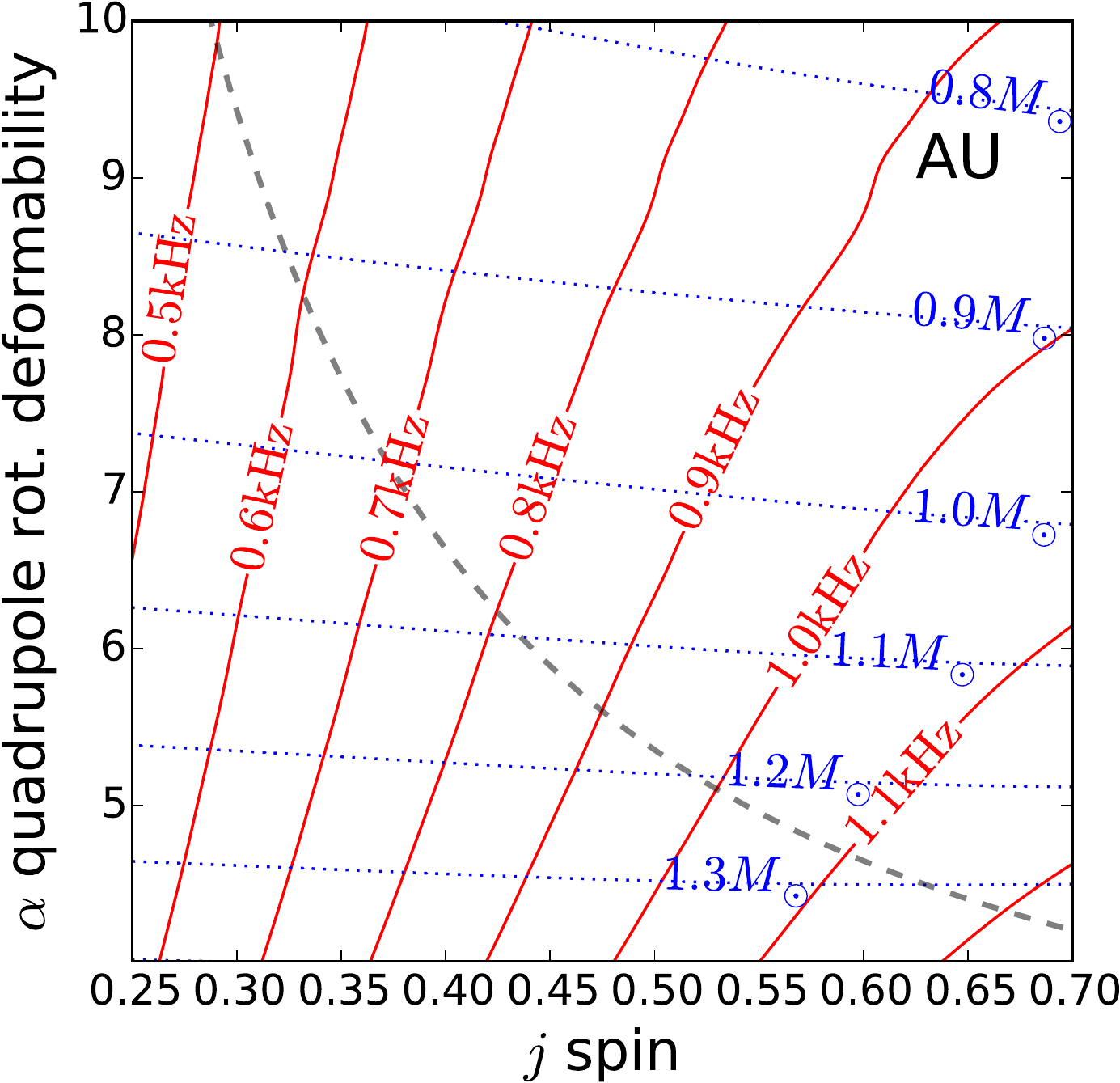} &
\includegraphics[width=0.49\columnwidth]{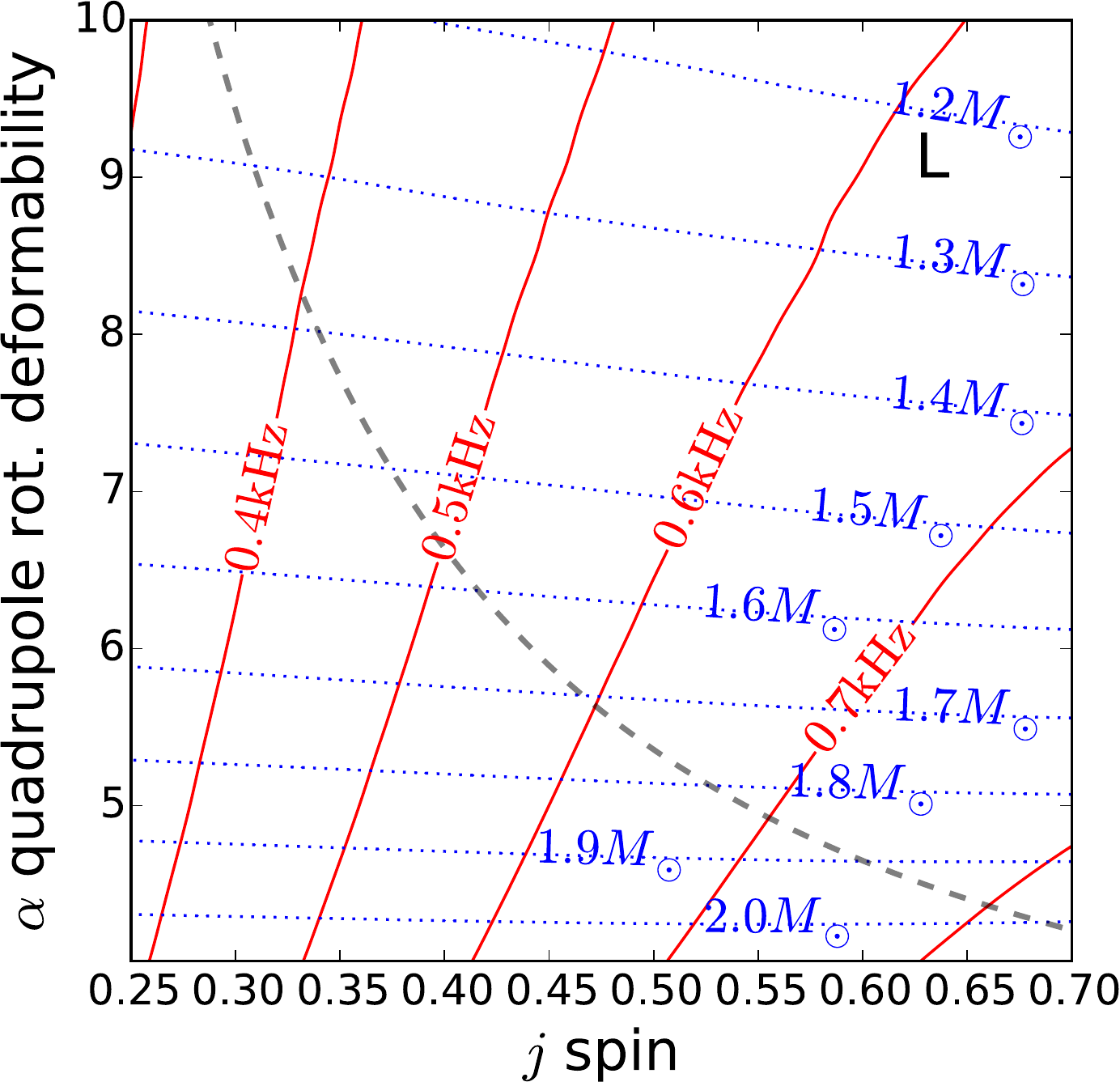}
\end{array}$
\end{center}
\caption{NS spin frequency (red-solid) and gravitational mass (blue-dotted) contours as a function of $j$ and $\alpha$ for various EOSs. In order of softest to stiffest, the EOSs considered are AU, SLy4, APR, and L. SLy4 and APR are both considered mid-range realistic EOSs, while AU and L are used here to represent the soft and stiff limits of realistic EOSs. The grey-dashed line is the same as in Figure \ref{fig:LT_peakparam} and shows the parameters where $r_{\rm peak} = r_{\rm NS}$. Higher spin-frequency, lower mass NSs tend to be above this line such that self-trapping of c-modes can occur. Stiffer EOSs yield a larger quadrupole rotational deformability, $\alpha$, for a given mass, and a higher dimensionless spin, $j$, for a given rotation frequency.} 
\label{fig:EOSparams}
\end{figure}

\section{Spin Frequency and Mass Bounds for c-Mode Self-Trapping}\label{sec:EOS}
In the previous section we determined the range of the $j$ and $\alpha$ spacetime parameters (Figure \ref{fig:LT_peakparam}) such that c-modes can become self-trapped. However, in order to convert these $j$ and $\alpha$ ranges into ranges of the more familiar NS properties of spin frequency and mass, an EOS must be selected. 

Here we examine NSs with four different standard EOSs: AU, SLy4, APR, and L, in order of increasing stiffness. We have chosen SLy4 and APR to represent mid-range realistic NS EOSs, while AU and L we use as proxies for the soft and stiff extremes for realistic EOSs. In Figure \ref{fig:EOSparams} we show NS mass and spin frequency contours as a function of $j$ and $\alpha$ for each of these EOSs. The mass contours are determined using interpolation of numerically generated RNS models \citep{StergioulasFriedman}. The models are the same used by \citet{Pappas:2013naa} (see Supplemental material of that paper for details) and cover the full range of NS parameters for each EOS. \citet{Pappas2015} showed in Appendix B that the quantity $M\times f_{\rm spin}$ can be given by a single fitting formula, for the range of realistic EOSs, as a function of $j$ and $\alpha$ that is accurate to within $2\%$ across this parameter range. The frequency contours of Figure \ref{fig:EOSparams} are determined using this fitting formula, and the interpolated masses. 

From these contours we see that to have self-trapping c-modes for mid-range EOSs (SLy4, and APR) we require 
fast spins ($f_{\rm spin} \go 0.6$ kHz), and low masses $M \lo 1.2 M_\odot$. Comparing AU and L we see that stiffer EOSs have higher $\alpha$ for a given mass, and higher $j$ for a given spin frequency, such that extremely stiff EOSs, like L, allow for self-trapping to occur for a much wider range of masses and spin frequencies. Extremely soft EOSs, like AU, require unrealistically low-mass and high-spin in order for a NS to have self-trapping c-modes. In general, the possible range of NS parameters for which c-modes can be observed will be different for different EOSs, providing the possibility to distinguish between them. 

\section{Discussion}

We have examined the diskoseismic corrugation modes for thin accretion disks using the relativistic free-particle characteristic frequencies, $\Omega$, $\Omega_\perp$, and $\kappa$,  in spacetimes with more prominent higher-order multipoles. Such spacetimes can arise due to spin-induced deformation of a NS contributing to the mass-quadrupole, spin-octupole, and higher-order RMMs. Expanding up to the lowest-order spin-octupole contribution we have determined the characteristic frequencies for these spacetimes in terms of the mass $M$, the spin parameter $j$, the quadrupole rotational deformability $\alpha$, and spin-octupole rotational deformability $\beta$. Using these frequencies, and the disk perturbation equations, we have shown that c-modes can become self-trapped in NS spacetimes (Figure \ref{fig:NSprop}) due to a turnover of the Lense-Thirring frequency $\Omega_{\rm LT} = \Omega - \Omega_\perp$, for sufficiently high values of the parameters $j$ and $\alpha$ (Figure \ref{fig:LT_peakparam}). 

The masses and spin-frequencies corresponding to these values of $j$ and $\alpha$ depend on the stiffness of the EOS. In Figure \ref{fig:EOSparams} we showed the various mass and spin-frequency ranges above which c-modes can become self-trapping. Mid-range realistic EOSs, like APR or SLy4, require a combination of very high spin frequencies ($\go 0.6$ kHz), and low masses ($\lo 1.2 M_\odot$) in order to reach this threshold. The fastest known millisecond pulsar, PSR J1748--2446ad, has spin frequency 716 Hz \citep{Hessels2006}, while the lowest well-constrained NS masses currently measured are  $\sim1.0 M_\odot$-$1.1 M_\odot$ \citep{Lattimer2012, Martinez2015}. Thus NS spin and mass combinations sufficient to induce c-mode self-trapping for mid-range EOSs are plausible but likely very rare. This, of course, ignores evolutionary considerations, e.g., that recycled millisecond pulsars require significant accretion in order to be spun up, which may raise the lowest plausible mass by up to $\sim 0.1 M_\odot$, depending on magnetic accretion torque history \citep{Alpar1982, Tauris2012}.
 
For stiffer EOSs, these mass and spin requirements are loosened. For the extremely stiff L EOS a $1.4 M_\odot$ NS with spin period $\go 400$Hz can have c-mode self-trapping. However, extremely soft EOSs, represented here by AU, require unrealistically high-spins and low-masses to induce c-mode self-trapping. 

If high-spin, low-mass accreting NS candidates can be identified, it may be possible to detect self-trapping c-modes, e.g., through the use of sufficiently high-resolution phase-resolved spectroscopy of the Fe-K$\alpha$ fluorescence line \citep{Miller2005, Tsang2013, Ingram2015}. The broadened Fe-K$\alpha$ line probes the structure of the inner part of the disk, such that the radial propagation region for the c-modes could be identified through the redshift range of the iron line variability at the c-mode frequency \citep{Tsang2013}. Such observations may be possible with new or upcoming X-ray missions such as LOFT \citep{LOFT} or Astro-H \citep{AstroH}.

The self-trapping of c-modes in NS accretion disk represents a qualitative difference in the behavior of the system due to the influence of the spacetime quadrupole. Identifying these self-trapped modes can then help to constrain the NS quadrupole, as well as the NS EOS through its rotational deformability. 

Finally, we point out that similar effects could also be observed for more ``exotic'' compact objects. The turnover of $\Omega_{\rm LT}$ would also be present for quark stars \citep{Gondeketal}, therefore self-trapping c-modes should be expected from these objects as well. Moreover, since quark stars can have smaller masses and radii than NSs as well as higher spin parameters \citep{Gourgoulhon1999}, the range of $j$ and $\alpha$ where such quark stars could develop self-trapping c-modes should be significantly larger. 

Compact objects in alternative theories of gravity could also be considered, as NSs and BHs in some of these theories, due to the presence of additional degrees of freedom, can develop additional ``hair'' that modify the spectrum of multipole moments and the geodesic frequencies \citep{Pappas2015a,Bertietal2015}. Self-trapping c-modes might be present in these theories for a quite different range of parameters, distinguishing them from the GR predictions. For example, a rotating BH of an alternative theory that deviates from Kerr by having a larger quadrupole would have self-trapping c-modes in contrast to the usual Kerr BH.

%

\section*{Acknowledgments}

D.T. was supported by the UMCP-Astronomy Center for Theory and Computation Prize Fellowship program. G.P. was supported by NSF CAREER Grant No.~PHY-1055103.
We thank L.C. Stein, E. Berti, C. Mingarelli, and M.C. Miller for useful discussion. 


\end{document}